\documentclass[twoside,fleqn]{article}
\usepackage{espcrc2}
\usepackage{epsfig}

\title{%
Centre vortices and their friends}

\author{P.W.~Stephenson%
  \address{Dipartimento di Fisica and INFN, Via Buonarroti 2,
    56127 Pisa, Italy}}

\begin{document}
\begin{abstract}
  We report two probes of centre vortices in pure
  SU(2).  First, we attempt to generate plaquette-size Z(2) vortices in a
  quasi-random way to compare the structure with those of projected centre
  vortices from the full theory; our prescription is, however, not
  completely random.
  Second, we test the effect of centre vortices on Wilson loops as a
  function of position.  This shows clearly that the major effect is
  the piercing of the Wilson loop regardless of the position of the
  vortex inside.
\end{abstract}

\maketitle

\section{INTRODUCTION}

One approach to QCD confinement is that of centre vortices, where
the key degrees of freedom are those in the centre of the gauge group,
Z($N$) in the case of SU($N$).  Numerical results for the case of SU(2)
(also treated here) show that this is fruitful.  The main technique is that
of centre projection~\cite{dDFGO97}: fixing to a gauge where the links are
as close as possible to a centre element, then projecting to that element,
leaving a lattice of Z(2) links; negative plaquettes are called P-vortices
and are interpreted as the source of confinement.  Here we examine two
related issues.

\section{RANDOM VORTICES}

A random gas of infinitely long vortices will cause linear confinement.
This is too simplistic, but maybe can teach us something: indeed it gives
about the right string tension from measured vortex densities.  Viewed in
four dimensions the vortices are defined by closed surfaces; confinement
survives only so long as this surface percolates throughout the space-time
manifold, and hence deconfinement may be due to loss of
percolation~\cite{ELRT99}.

This has all been argued from the point of view of taking SU(2) and
reducing the degrees of freedom to the bare essentials.  Here we shall
attempt the opposite: to construct an (approximately) random vortex picture.
Truly random vortices are difficult because of the strong coupling of
adjacent plaquettes via the links, even with no gauge coupling present.
Our lattice and observables are as in the projected Z(2) theory.  We use
the following procedure:

\begin{itemize}
 \item Create a random set of links, either $\pm1$ with 50\%
  probability (`random start') or set to unity (`frozen start').

 \item Let $v = \textrm{density of negative plaquettes}$ (corresponding to
  vortices); initially $v\sim0.5$ or $v\equiv0$.  Pick a target
  $v=v_T$ chosen to correspond to the mean density of P-vortices in SU(2).
  At $\beta=2.3$, $v_T\approx0.0945$; at $\beta=2.4$, $v_t\approx0.0602$.
  
 \item Pick a link at random in the lattice.  Flip the sign of this link
  either (i) if it does not alter $v$ or (ii) if it alters $v$ towards
  $v_T$.

 \item Continue until $v_T$ is achieved.  Because of condition (i) it is
  useful to attempt to flip links already considered.
  In the case of the frozen start, we have tried further to make the
  vortices independent by making sets of flips which do not affect the
  overall vortex density.

 \item Generate many configurations of this sort and analyse them as a
  Monte Carlo sample.  Note that here there is no Markov process, and hence
  no fluctuating action; in a sense our ensemble is microcanonical.
\end{itemize}

There is a bias in this proceedure because we flip links attached to sets
of plaquettes predominantly of one sign, hence our vortices are not truly
random.  We could instead have chosen the target $v$ to correspond to the
SU(2) string tension on the assumption of truly random vortices.  Our
actual choice reflects a desire to look at the cluster properties of
vortices.

\subsection{Results}

\begin{table}
\begin{tabular}{r|lll}
  \hline
                    & $\beta=2.3$           & Random           & Frozen \\
  \hline
  (a)               & $0.0945(6)$           & $0.0945$       & $0.0945$ \\
  (b)               & $0.0727(9)$           & $0.01060(11)$  & $0.9876(2)$ \\
  (c)               & $10890(60)$           & $11631(1)$     & $146(2)$ \\
  (d)               & $0.1362(2)$           & $0.189$       & $0.189$ \\
  (e)               & $0.145(7)$            & $0.41(3)$      & $0$ \\
  \hline
  \hline
                    & $\beta=2.4$ \\
  \hline
  (a)               & $0.06015(6)$          & $0.0602$      & $0.0602$ \\
  (b)               & $0.1125(12)$          & $0.00494(6)$  & $0.99742(3)$ \\
  (c)               & $20600(100)$          & $23553(1)$    & $61.0(6)$ \\
  (d)               & $0.0708(11)$          & $0.12$      & $0.12$ \\
  (e)               & $0.079(7)$            & $0.164(3)$    & 0 \\
  \hline
\end{tabular}
  \smallskip
\caption{Comparsion between SU(2) and quasi-random vortices. The columns
  show SU(2), and quasi-random vortices starting from random or frozen
  links;  the rows
  are (a) vortex density $v$ (fixed for the quasi-random vortices), (b)
  fraction of P-plaquettes \textit{not} in the largest cluster, (c) the
  number of P-plaquettes in the largest cluster,
  (d) the string tension based on large scale simulations
  of full SU(2)~\protect\cite{BSS} or purely random vortices, and (e) that
  actually measured from vortices, see text.}
\label{fig:qrres}
\vskip -4ex
\end{table}

Fig.~\ref{fig:qrres} shows results on bulk lattices, $12^4$ for $\beta=2.3$
and $16^4$ for $\beta=2.4$.  The string tension is shown both for the two
ideal cases (from a large scale run in full SU(2) and for fully random
vortices) and as measured from vortices.  In the quasi-random case with the
random start, Creutz ratios show a string tension which for small loops
lies near the expected value ($2v$) but which increases for larger loops.
The results shown are from a full potential calculation where this increase
tends to level out, although with some curvature, giving a rather larger
string tension; the form fit to is necessarily somewhat ad hoc and here we
have included a quadratic part.  Furthermore, in the frozen start the
vortices lack confinement and hence show in effect a repulsion.  These are
sizeable effects; a more truly random method will be needed for a more
realistic comparison.  An effective action would also presumably
help~\cite{ER99}.

Nonetheless, we examine cluster properties by methods similar to
ref.~\cite{BFGO99}, dividing vortices into two clusters where the surfaces
touch only along an edge.  This difference between touching and joining is
a lattice effect which makes a noticeable impact --- almost tripling the
number of vortices not in the largest (percolating) cluster for the case of
SU(2) with $\beta=2.3$ with the random start, and increasing the largest
cluster size dramatically for the frozen start.  Of course we would prefer
to detect vortices directly with their physical size.

\subsection{The deconfining transition}

We have also examined a lattice in the deconfined phase, using Polyakov
loops $L$ as the order parameter, although it is maybe unlikely that
homogeneous random vortices alone can be sufficient to explain
deconfinement.  The lattice results show that $\langle\left|
L\right|\rangle$ goes to 1 for small vortex density, but this is expected
simply due to the fact that neighbouring loops are effectively Wilson loops
with an area equal to the finite temperature extent of the lattice, and
hence correlated by the vanishing string tension.  There is no sign of a
phase transition, nor finite size scaling behaviour.  It may well be
important to have the vortex surface orientated predominantly parallel to,
and hence not piercing, temporal Wilson loops; it is not clear such an
effect can come from just the Z(2) degrees of freedom.

\section{PROBING WILSON LOOPS}

The plaquette-sized P-vortices are expected to have a topological effect on
Wilson loops, depending only on whether a vortex pierces the loop.  We
investigate this by looking at the correlations between P-vortices and
Wilson loops.

\begin{figure}
  \begin{center}
    \psfig{file=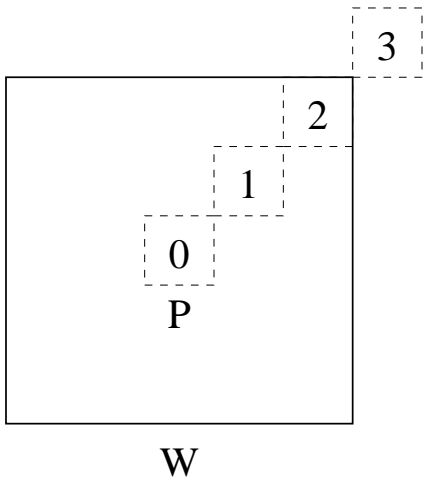,width=2.5in,angle=270}
  \end{center}
  \vskip -5ex
  \caption{Placement of a plaquette-sized probe in a Wilson loop, showing
    the diagonal distance $r$; for a loop size $L$ with $L$ odd, $r =
    (L+1)/2$ is just outside the loop.}
  \label{fig:pwl}
  \vskip -2ex
\end{figure}

Our method is the following (fig.~\ref{fig:pwl}).  We take a plaquette $P$
on the centre-projected lattice within a wilson loop $W$, a certain
distance from the centre of the loop.  For present purposes we shall simply
take the distance $r$ to be the number of plaquettes diagonally from the
centre of the loop, as in the diagram.  If $P=1$, we ignore $W$ and pass on
to the next one; if $P=-1$ we examine the value of $W$.  After sampling over
many configurations, we can form an average $\langle W_{P=-1}(r)\rangle$.
Note that in examining $P$ we take no account whatsoever of other centre
plaquettes inside (or outside) $W$; the effect is purely the correlation
between the Wilson loop and a centre vortex at the given position, whether
or not the loop is pierced by other vortices.  To achieve sufficiently
large correlations we are restricted to loops of sizes that have
$\mathcal{O}(1)$ vortices inside. Clearly, if there is no correlation,
$\langle W_{P=-1}(r)\rangle=\langle W\rangle$.  As a control, we have
performed the same experiment replacing $P$ with the sign of a gauge
plaquette $G$ located in the same place.

\begin{figure}
  \vskip -2ex
  \begin{center}
    \psfig{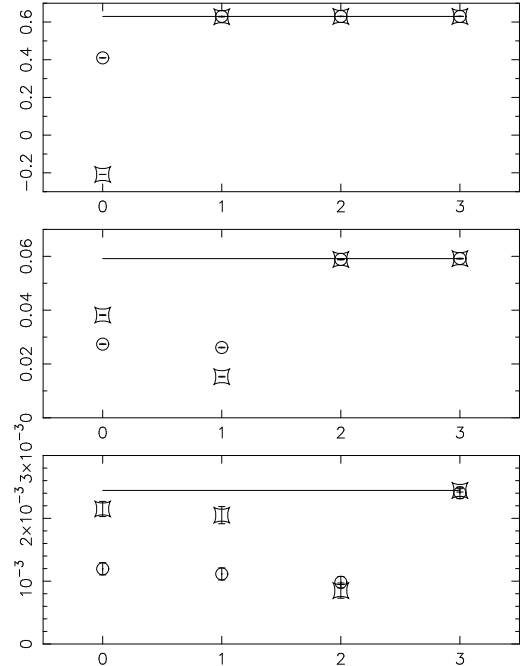}
  \end{center}
  \vskip -5ex
  \caption{Probes $W_{P=-1}$ (circles) and $W_{G=-1}$ (squashy squares) for
    $\beta=2.4$, $16^4$ lattice, for (top) $1\times1$, $3\times3$,
    $5\times5$ (bottom) loops.}
  \label{fig:pwlres}
  \vskip -3ex
\end{figure}

The results (fig.~\ref{fig:pwlres}) show that $\langle W_{P=-1}(r)\rangle$
is rather flat inside the loop, but with a significant correlation.  In
contrast, the values of $\langle W_{G=-1}(r)\rangle$ vary much more widely
over the inside of the loop.  This is a sign that the dominant effect of
the vortex is given by whether or not it pierces the loop, regardless of
where it does so, an effect not expected and not shown by the sign of the
full gauge plaquette.  Both probes become uncorrelated very quickly when
outside the loops.  For gauge plaquettes this can be understood from strong
coupling; such plaquettes only appear in quite high order.  For P-plaquttes
the natural interpretation is that vortices not piercing the Wilson loop
have no effect on it.  However, if the vortices really correspond to
extended physical objects, it is not clear why the change from inside to
outside should be so sharp; this raises questions about the size of the
vortex core.

\end{document}